\documentclass[conference]{IEEEtran}
\pdfoutput=1
\IEEEoverridecommandlockouts
\usepackage{amsmath,amssymb,amsfonts}
\usepackage{algorithmic}
\usepackage{graphicx}
\usepackage{textcomp}
\usepackage{xcolor}
\def\BibTeX{{\rm B\kern-.05em{\sc i\kern-.025em b}\kern-.08em
    T\kern-.1667em\lower.7ex\hbox{E}\kern-.125emX}}

\usepackage{tikz}
\usetikzlibrary{shapes,positioning}
\tikzset{%
  io/.style={minimum width=3cm, minimum height=0.5cm, text centered},
  process/.style={rectangle, minimum width=3cm, minimum height=1cm, text centered, draw=black},
  decision/.style={diamond, minimum width=3cm, minimum height=1cm, text centered, draw=black,aspect=4},
  arrow/.style={thick,->,>=stealth},
  sender/.style={rectangle, draw, fill=blue!20, text width=9.5cm, text justified, minimum height=1.0cm, minimum width=5.0cm, rounded corners, anchor=west,xshift=-2cm},
  receiver/.style={rectangle, draw, fill=green!20, text width=9.cm, text justified, minimum height=1.0cm, minimum width=5.0cm, rounded corners, anchor=east,xshift=2cm}
}

\usepackage{hyperref}
\usepackage{cleveref}
\usepackage{listings}
\definecolor{codegreen}{rgb}{0,0.6,0}
\definecolor{codegray}{rgb}{0.5,0.5,0.5}
\definecolor{codepurple}{rgb}{0,0.58,0.82}
\definecolor{backcolour}{rgb}{0.95,0.95,0.92}
\lstdefinestyle{mystyle}{
    backgroundcolor=\color{backcolour},   
    language=java,
    commentstyle=\color{codegreen},
    keywordstyle=\color{codepurple},
    numberstyle=\tiny\color{codegray},
    stringstyle=\color{codepurple},
    basewidth=.5em,
    basicstyle=\ttfamily,
    breakatwhitespace=false,         
    breaklines=true,                 
    captionpos=b,                    
    keepspaces=true,                 
    showspaces=false,                
    showstringspaces=false,
    showtabs=false,                  
    tabsize=4
}
\crefname{lstlisting}{listing}{listings}
\Crefname{lstlisting}{Listing}{Listings}

\usepackage{xspace}
\newcommand{\KeY}{Ke\kern-.1emY\xspace}

\usepackage[style=authoryear-icomp,maxcitenames=2,backend=biber,style=lncs,sortcites=true,minbibnames=2,maxbibnames=3]{biblatex}
\begin{document}

\title{
Next \emph{Steps} in LLM-Supported Java Verification 
\thanks{This work was supported by funding from the pilot program Core-Informatics of the Helmholtz Association (HGF).}
}

\author{\IEEEauthorblockN{Samuel Teuber}
\IEEEauthorblockA{\textit{Karlsruhe Institute of Technology}\\
Karlsruhe, Germany \\
teuber@kit.edu}
\and
\IEEEauthorblockN{Bernhard Beckert}
\IEEEauthorblockA{\textit{Karlsruhe Institute of Technology}\\
Karlsruhe, Germany \\
beckert@kit.edu}
}

\maketitle

\begin{abstract}
Recent work has shown that Large Language Models (LLMs) are not only a suitable tool for code generation but also capable of generating annotation-based code specifications.
Scaling these methodologies may allow us to deduce provable correctness guarantees for large-scale software systems.
In comparison to other LLM tasks, the application field of deductive verification has the notable advantage of providing a rigorous toolset to check LLM-generated solutions.
This short paper provides early results on how this rigorous toolset can be used to \emph{reliably} elicit correct specification annotations from an \emph{unreliable} LLM oracle.
\end{abstract}


\section{Introduction}
\looseness=-1
Along with the pervasive success of Large Language Models~\cite{DBLP:conf/nips/BrownMRSKDNSSAA20} (LLMs) across many application domains, a steadily growing body of research is exploring the usage of LLMs for the synthesis of specification annotations based on a program's source code~\cite{kamath2023finding,chakraborty2024ranking,10.1007/978-3-031-57259-3_13,wu2024lemur,DBLP:journals/corr/abs-2311-03739,pmlr-v202-pei23a,LATHOUWERS2024111972,granberry2024specify,10.1007/978-3-031-65112-0_7,Beckert24}.
The appeal of using LLMs in this context comes from the ``NP-like'' imbalance in the problem formulation:
While it is, in general, arbitrarily difficult to generate a correct, \emph{useful} specification for a given piece of code, auto-active verifiers~\cite{leino2010usable}, in principle, allow us to \emph{verify the correctness} of a given specification.
The combination of LLM-based specification synthesis and deductive specification verification has now been validated in multiple case studies and has even been applied for real-world programming languages such as Java and JML~\cite{JML-Ref-Manual} using the verification tool \KeY{}~\cite{DBLP:series/lncs/AhrendtG16} (see \cite{Beckert24}).

In this light, LLM-based specification generation, as a canonical example of Intersymbolic AI~\cite{Pla24}, has the potential to usher in a ``golden age'' of deductive verification in which the average software engineer can finally apply rigorous program analysis techniques to their large-scale software systems and prove system correctness on their own.
Generating a method's top-level specification still requires human intervention to ensure the specification matches the requirements, but it has the potential to ease the workload to achieve provable correctness.
This work focuses on the other promising application field, namely auxiliary specification generation.
Here, the top-level specification is still provided by a human and the LLM works under the supervision of an \emph{automated} verifier which checks whether the LLM's annotations are sufficient to prove the top-level specification.

\looseness=-1
Unfortunately, we are still far from reaching the golden age. 
One reason for this is a lack of techniques to \emph{reliably} elicit annotations from an LLM.
As the number of required annotations grows with the size of the analyzed software system, a lack of reliability quickly hinders our ability to scale the approach.

\paragraph*{Contribution}
\looseness=-1
In this work, we present early results on exploring different prompting paradigms for the reliable generation of JML contracts via LLMs.
In particular, we analyze to what degree feedback from the verifier is currently helping the specification generation.
Finally, we discuss the necessary next steps to scale our approach to larger-scale systems.
The detailed prompts used for our experiments can be found in \Cref{appendix}.

\section{LLMs for Auxiliary Specification Generation}
This work considers the specification of Java programs using the Java Modelling Language (JML)~\cite{beckertModularVerificationJML2020}.
JML allows the contract-based specification of Java methods and classes via JavaDoc-like comments and is a supported specification language of multiple Java verifiers~\cite{DBLP:series/lncs/AhrendtG16,beckertModularVerificationJML2020,DBLP:journals/corr/Cok14}.
In this work, we made use of the \KeY{} tool~\cite{DBLP:series/lncs/AhrendtG16}, which deductively verifies that a Java program adheres to a given contract-based JML specification.
\KeY{} comes with a built-in proof search strategy that allows automated verification and supports modular verification:
Using the abstractions provided by method contracts, specifications of larger software systems can be decomposed into verification tasks for individual methods.

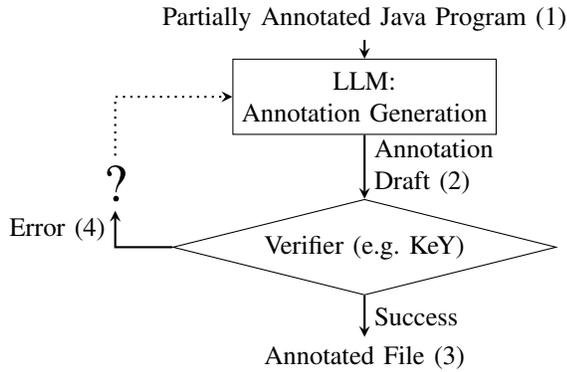
\begin{figure}
    \centering
    \begin{tikzpicture}[node distance=0.05cm]
        
        \node (jmlfile) [io,anchor=center,align=center] {Partially Annotated Java Program (1)};
    
        \node (gpt) [process, below=of jmlfile,align=center,yshift=-0.2cm] {LLM:\\Annotation Generation};
        \draw[arrow] (jmlfile) -- (gpt);

        \node (questionmark) [io,anchor=center,align=center,left=of gpt,yshift=-1.15cm] {\huge ?};
    
        \node (key) [decision,below=of gpt,yshift=-0.8cm,align=center] {Verifier (e.g. \KeY{})};
    
        \draw[arrow] (gpt.south) -- node[pos=0.5,anchor=west,align=left] {Annotation\\Draft (2)} (key.north);
    
        \node (output) [io, below=of key,yshift=-0.5cm] {Annotated File (3)};
    
        \draw[arrow] (key.south) -- node[anchor=west] {Success} (output.north);
    
        \draw[arrow] (key.west)  -| node[anchor=east,pos=0.75]{Error (4)} (questionmark.south);

        \draw[arrow, dotted] (questionmark.north)  |- (gpt.west);
    \end{tikzpicture}
    \caption{Setup of our LLM integration for \KeY{}: A partially JML-annotated Java file is handed to an LLM that generates additional annotations for the file. Subsequently, the file with extended annotation is handed to a verifier (in this instance \KeY{}) which attempts to verify the correctness of the code w.r.t. the given annotations. This work explores strategies to recover from errors when the verifier cannot prove the specification draft of the LLM.}
    \label{fig:overall_approach}
\end{figure}

\paragraph*{LLM-based synthesis of JML annotations}
\looseness=-1
In prior work~\cite{Beckert24}, we proposed an architecture for combining (deductive) verification tools such as \KeY{}~\cite{DBLP:series/lncs/AhrendtG16} with LLM-based specification generation and evaluated our approach on a set of Java programs.
In our approach, outlined in \Cref{fig:overall_approach}, a partially annotated Java Program~(1) is handed to an LLM with instructions to complete the annotation.
Subsequently, the annotations generated by the LLM are added to the program code, and the \emph{Annotation Draft}~(2) is handed to a verification tool, e.g.\ the Java verification tool \KeY{}.
If \KeY\ successfully verifies the correctness of the Annotation Draft, we return the annotated file~(3).
If verification fails, we obtain a description of the error~(4) from the verifier and now require a strategy to recover from this error -- we will further discuss this part of the approach below.

\looseness=-1
Depending on the application context, the guarantees provided by our approach can vary:
If the Java Program~(1) contains no prior annotations and the LLM extends the program by a verifiable annotation for a top-level method, we do not have any guarantee beyond internal consistency between the generated code and its specification.
On the other hand, if the partially annotated Java Program contains a top-level specification and the LLM only complements the annotation by auxiliary specifications such as loop invariants and/or method contracts for submethods, successful verification guarantees that the (human-provided) top-level specification is satisfied by the implementation.

\looseness=-1
We believe the latter use case is particularly promising as it only delegates work to the LLM that cannot compromise soundness and is thus resistant to an LLM's errors:
If the generated annotations admit verification, we know that the user-given top-level specification is satisfied due to the rigorous deduction inside \KeY{}.
Hence, we will focus on the generation of auxiliary annotations for a given top-level specification.
To focus our study on the LLM's capabilities,
we consider the case where the Java+JML file has \emph{one gap} in the specification that the LLM has to complete (e.g. one missing loop invariant).

\begin{lstlisting}[caption={Callee method \texttt{g} lacks an annotation},label=lst:submethod]
//@ ensures \result == -2*x;
int f(int x) { return g(-x); }
int g(int x) { return x+x; }
\end{lstlisting}

\paragraph*{Example}
\looseness=-1
Consider the example in \Cref{lst:submethod}: In this instance, we have a JML annotation for the method \texttt{f}.
The ensures clause defines a contract for~\texttt{f} stating that its return value should be \texttt{-2*x}.
To prove this contract in a modular way, we require an additional contract for method \texttt{g}, which is called by method \texttt{f}.
This contract could be provided by an LLM:
If the LLM manages to synthesize a contract for \texttt{g} such that \KeY{} can prove the correctness of both methods (w.r.t.\ the contract in \Cref{lst:submethod} resp.\ the contract generated by an LLM), we know that the top-level specification of~\texttt{f} is satisfied.
This methodology of annotation-based automated verification is sometimes called \emph{auto-active} verification~\cite{leino2010usable}.

\begin{figure}
    \centering
    \includegraphics[width=1.0\linewidth,trim={0cm 0.25cm 0 0.25cm},clip]{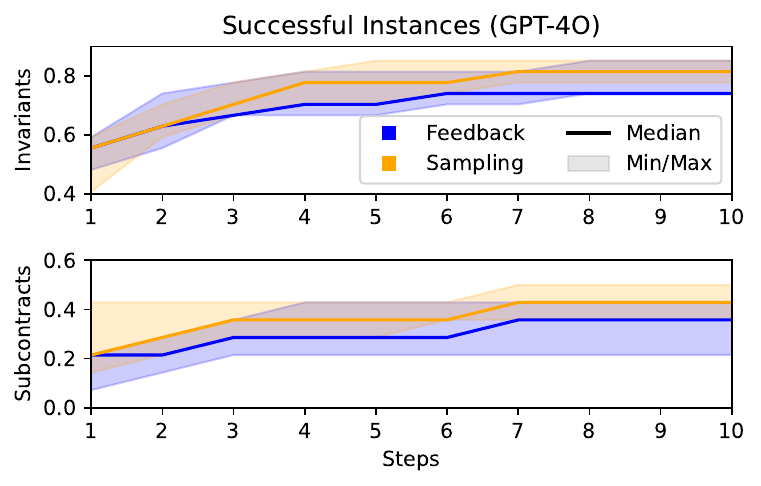}
    \caption{Increase of Success Rate for invariant (top) resp.\ subcontract (bottom) synthesis using GPT 4O for increasing step rate: Shaded areas represent minimal/maximal success rates over 5 runs, the solid line represents the median.}
    \label{fig:success_steps}
\end{figure}
\section{Error Recovery}
\looseness=-1
In prior work~\cite{Beckert24}, we used rudimentary error messages from \KeY{} to generate feedback for the LLM.
Here, we can distinguish two kinds of failures: syntactical errors in the generated specification (e.g., missing semicolons, wrong use of variables, etc.)\ and semantical errors (e.g., a postcondition that is not satisfied by the method's implementation).
For syntactical errors, we provide the LLM with the parser's error message;
for semantical errors, the LLM receives the (\KeY{}-generated) labels of open proof branches.
These labels, describe which part of the proof failed, e.g., the part where \KeY\ tried to prove that a loop invariant is initially valid ("Invariant Initially Valid") or that the post-condition holds when the value of an array is not null("Normal Execution (\_a != null)").

\looseness=-1
However, this raises the question of whether the provided feedback is helpful for the LLM in correcting the specification.
As a simple baseline comparison, we can use a \emph{sampling-based} approach:
Instead of providing feedback to the LLM and continuing the same conversation string, we can restart the (probabilistic) process from scratch whenever an LLM-generated specification cannot be verified.
To this end, we reran the experiments from prior work for the synthesis of loop invariants and subcontracts with the current version of GPT~4O and compared the previous feedback-based approach to a sampling-based approach, which does not provide \KeY{}'s feedback to the LLM.
As a benchmark set, we used 27 tasks for JML invariant synthesis and 14 tasks for the specification of called methods.
The benchmarks cover a wide range of JML features including array usage, quantifiers, assignable clauses, reference to pre-execution values, etc.\ (see \cite[Table 1]{Beckert24}).
For each instance, we sampled (i.e., restarted the process) up to 10 times.
\Cref{fig:success_steps} provides a first overview of these results:
Here, we see how the rate of success increases with the number of feedback steps resp.\ sampled solutions for invariant and subcontract synthesis.

\looseness=-1
On the one hand, the results show that an iterative process with feedback increases the likelihood of success.
This is a positive result because it means we are not in a situation where GPT 4O either is capable of generating a solution or consistently fails across all attempts.
On the other hand, these results indicate that the current feedback provided by \KeY{} does not significantly improve GPT 4O's ability to generate correct JML contracts:
Because, even without any feedback, GPT 4O achieves similar success rates and sometimes even outperforms the feedback-based approach.
We suspect the latter result can be partially explained by the observation that GPT~4O sometimes gets ``stuck'' on a wrong solution and only performs marginal changes to the previous, wrong solution instead of pivoting to a correct solution that would require more changes.

\begin{figure}
    \centering
    \includegraphics[width=\linewidth,trim={0cm 0.25cm 0 0.25cm},clip]{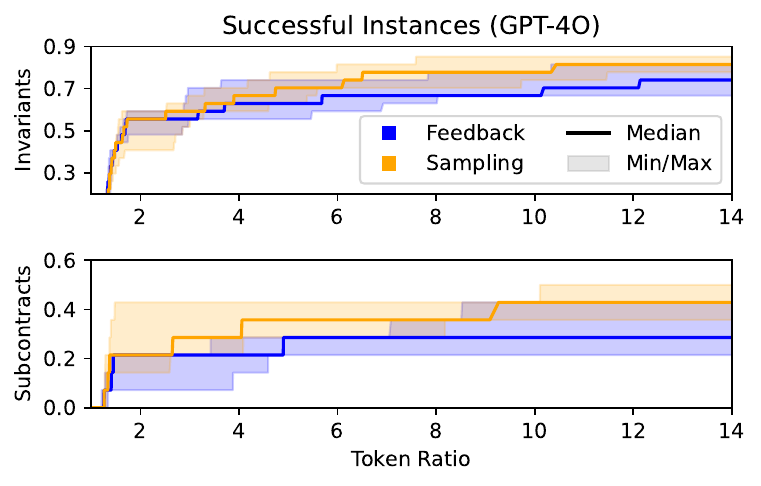}
    \caption{Increase of Success Rate for invariant (top) resp.\ subcontract (bottom) synthesis using GPT~4O for increasing token ratio: Shaded areas represent minimal/maximal success rates over 5 runs, the solid line represents the median.}
    \label{fig:success_ratio}
\end{figure}

\paragraph*{Token Efficiency}
\looseness=-1
Another perspective on the two competing approaches (sampling and feedback) is token efficiency:
LLM text generation is typically priced in \emph{tokens}, which are the elementary components of a text that an LLM processes to perform its predictions.
For this evaluation, it is important to note that different problem instances require different amounts of tokens:
For example, assuming fixed output length, generating a specification for a method with 100~lines of code will necessarily require more tokens than generating code for a method with 10~lines of code (the token count comprises both read input tokens and output tokens written by the LLM).
Hence, we divide the number of tokens used to find a solution by the number of tokens in the initial input query to the LLM and call this metric the \emph{token ratio}.
The success rate w.r.t. token ratio is plotted in \Cref{fig:success_ratio}.

\looseness=-1
As a first observation, we note that sampling and feedback-based approaches fundamentally differ in their token efficiency:
Feedback-based approaches input the entire previous conversation back into the LLM for the generation of the next answer.
Thus, if we initially input $I$ tokens, the LLM generates $O$ tokens on each turn and each feedback response on our side comprises $R$ tokens, then the first iteration costs $I+O$ tokens, the second iteration costs $I+O+R+O$ tokens and the $n$th iteration costs $I+\left(n-1\right)\left(O+R\right)+O$ tokens.
In contrast, due to the lack of feedback, each sampling iteration only consumes $I+O$ tokens.
In total for $n$ iterations, we have token costs of $n(I+O)$ for sampling vs.\ $n(I+O)+\frac{n(n-1)}{2}(O+R)$ for feedback-based techniques.
Effectively, sampling grows linear in the number of steps while feedback-based techniques have a quadratic term.

Just as we previously plotted the success rate for a growing number of feedback/sampling steps, we can now plot the success rate for a growing allocated token ratio.
The results of this experiment are based on the same dataset as the previous experiment and consider 5 runs.
We evaluated our data up to token ratio 14 and the results can be found in \Cref{fig:success_ratio}.
In this perspective, sampling early on has a significant advantage over the feedback-based approach for subcontract generation and we also see a weak indication that sampling might be superior for invariant generation.

\paragraph*{A Mixed Approach}
Based on the observations from \Cref{fig:success_steps}, we hypothesize that the feedback-based approach eventually stagnates.
Moreover, the feedback-based approach cannot be scaled to arbitrary step sizes due to the size of the LLM's context window which only admits limited conversation length.
A logical step to scale our approach to even more attempts is thus to \emph{mix} the sampling-based approach with the feedback-based approach.
To this end, we can use the data gathered in our previous experiments to compare the success rate of 50~samples with the success rate of 5~samples each of which has a 10-step feedback iteration.
In both cases, up to 50 solutions are evaluated.
The results of the mixed approach can be observed in \Cref{tab:mixed}.
Importantly, there is an overlap between the unsolved cases: the mixed approach solves all benchmarks solved by sampling and two additional ones.
For invariants, the additional solved benchmark is a loop invariant for array reversal.
This invariant is found by GPT~4O at the 8th feedback step (in two out of five runs).
For subcontracts, the instance only found by the mixed approach is a contract for copying a range of an array.
However, that contract was found in the first iteration, i.e., without any feedback.

\begin{table}[t]
    \centering
    \caption{Results for mixed approach: 50 independent samples vs.\ sampling 5 rounds with 10-step feedback.}
    \begin{tabular}{l|l|l}
        \textbf{Category} & \textbf{Strategy} & \textbf{Success Rate}\\\hline\hline
        Invariant Generation & Pure Sampling & 23/27\\ \cline{2-3} 
        & Mixed & \textbf{24/27}\\ \hline 
        Subcontracts & Pure Sampling & 7/14\\ \cline{2-3}
        & Mixed & \textbf{8/14}
    \end{tabular}
    \label{tab:mixed}
\end{table}

\paragraph*{Results}
\looseness=-1
As a positive result, our experiments provide evidence that iterative approaches (whether feedback or sampling-based) improve the LLM's success rate for annotation generation.
This insight suggests that more refined, iterative strategies could lead to even stronger results which warrants further exploration.
Our observation also strengthens the case for an intersymbolic approach that couples LLM generation with verification as humans cannot be expected to inspect the LLM's many solutions candidates.
We also note that an approach solving approx. 85\% of JML invariant synthesis problems and 57\% of submethod contract problems is unquestionably promising.

\paragraph*{Limitations}
\looseness=-1
The shaded areas in \Cref{fig:success_steps,fig:success_ratio} indicate that our results are not yet statistically significant.
Although our benchmark contains a large number of relevant Java and JML features~\cite[Table 1]{Beckert24}, we require a larger set of representative instances to validate our hypotheses.
Moreover, our results only suggest that the \emph{current} feedback approach is insufficient to improve efficiency over sampling-based approaches.
This result could change through further prompt engineering which, again, requires a larger benchmark set. 
Finally, further research is necessary to extend our approach to larger gaps in the auxiliary specification (e.g. multiple missing loop invariants)

\section{Discussion}
This work presents our current methodology for evaluating different LLM prompting strategies for error recovery and presents some first evaluation results on error recovery in LLM-based JML synthesis.
We found that an iterative approach indeed improves the success rate which strengthens the argument for our verification-based intersymbolic approach.
For future work, we hope to repeat and extend the experiments presented using larger benchmark datasets.
More generally, we hope to extend our work in the following directions:

\paragraph*{Sampling vs. Feedback}
Unlike many other domains,
specification generation has the advantage of providing a vast set of tools to check the correctness of solutions.
In principle, we could thus ``simply'' generate a large number of solutions and find the needle in the haystack, i.e., the one specification that is consistent with our partially annotated input file.
There are of course limitations to this approach:
When our code has multiple annotation gaps, it may be necessary to fill all of them before verification, which quickly becomes work-intensive if an individual solution by the LLM has a low chance of being correct.
Hence, moving up the left end of the curve in \Cref{fig:success_steps,fig:success_ratio} is a highly desirable objective: We want to obtain successful specifications within the first attempts as often as possible.
We hope that an evaluation w.r.t.\ a recently released dataset~\cite{10.1145/3689484.3690738} for JML annotation may help us better understand the advantages and drawbacks of different prompting techniques in this context.

Future work will also focus on deriving textual representations of failed proof attempts that are more effective as feedback to the LLM.
How this information can be extracted from \KeY{}'s proof tree is a nontrivial question.

\paragraph*{Beyond Verification}
For the quick evaluation of specification candidates, 
we plan to explore additional checks using, e.g., testing/fuzzing before attempting verification.
Counterexamples could then also be used to devise feedback to the LLM.

\paragraph*{Beyond Filling Gaps}
Our experiments evaluated the ability of LLMs to fill the \emph{one} gap in a partially JML-annotated Java file.
As a next step, we want to devise a strategy that generates \emph{all} specifications to prove a given top-level specification (as a first step, by generating all necessary submethod and loop-invariant specifications).
This could greatly increase the practicability of the approach.

\printbibliography

\clearpage
\appendix
\subsection{LLM Prompts}
\label{appendix}
Below we provide examples for the prompts used to elicit specifications from the LLM.
\subsubsection{Invariant Generation}
We invoke GPT with the following system message:
\begin{lstlisting}[language={}]
You are an assistant for JML annotation.
In the first message of this conversation, you are provided with a Java class with partial JML annotation.
Additionally, you are provided with natural language instructions that describe the task to be completed.
JML annotation is a complicated task, but you are very capable and a perfect fit for this job.
First, think step by step: What does the code do? What is the context in which the code is executed? What variables are relevant?
Draft a behavioral description in natural language.
Subsequently, translate this description into JML annotation.
You should only provide the JML annotation and not the Java code.
We will then use the KeY verification system to check if the JML annotation is correct.
If the program verification fails, you will be provided with information about the failure and asked to correct the JML annotation.
Always add the JML keyword `normal_behavior` to the contract -- this guarantees that no exceptions are being thrown.
Your answers should always have the following format where the the <X> is substituted by the JML annotation suggested by you:
<your natural language reasoning>
```
/*@ <X>
*/
```
\end{lstlisting}
Subsequently, we prompt GPT with the following prompt to generate a loop invariant:
\begin{lstlisting}[language={}]
Given the following Java class:
'''
<partially annotated file>
'''
Please provide a loop invariant for the loop construct with the comment `//Add invariant here` of the method '<method name>'
Beware this annotation has to be a loop invariant.A loop invariant typically has the following structure:
```
/*@ loop_invariant ...;
  @ decreases ...;
  @ assignable ...;
@*/
```
\end{lstlisting}
For feedback-based approaches, we distinguish between syntactical and semantical errors.
For syntactical errors, we use the following prompt to continue the conversation:
\begin{lstlisting}[language={}]
The provided code is not valid JML. Please try again and make sure to provide a valid JML loop invariant.
This might describe the reason why change is required:
Beware that the error message may contain variable names from inside the proof system that are not available in the original code.
Such variable names must not in your JML answer; only use variable names from the original code.
<parser error>
\end{lstlisting}
An example for a parser error message could be:
\begin{lstlisting}[language={}]
Error during JML parsing: Failed to parse JML fragment: Encountered unexpected token: \"(\" \"(\"
    at line 3, column 7.

Was expecting one of: <list of possible options>
\end{lstlisting}

For semantical errors we use the following prompt:
\begin{lstlisting}[language={}]
The provided JML does not solve the task. The KeY verification told me that the JML is wrong because some proof goals were not closed:
Beware that the error message may contain variable names from inside the proof system that are not available in the original code.
Such variable names must not in your JML answer; only use variable names from the original code.
During verification, the following proof branches could not be closed:
<proof branch labels>
Please fix the JML loop invariant.
\end{lstlisting}
Examples for a proof branch label could be:
\begin{lstlisting}[language={}]
Normal Execution (_a != null)
\end{lstlisting}
\begin{lstlisting}[language={}]
Body Preserves Invariant
\end{lstlisting}
\begin{lstlisting}[language={}]
Use Case
\end{lstlisting}
\clearpage
\subsubsection{Submethod Contract Generation}
We invoke GPT with the following system message:
\begin{lstlisting}[language={}]
You are an assistant for JML annotation.
In the first message of this conversation, you are provided with a Java class with partial JML annotation.
Additionally, you are provided with natural language instructions that describe the task to be completed.
JML annotation is a complicated task, but you are very capable and a perfect fit for this job.
First, think step by step: What does the code do? What is the context in which the code is executed? What variables are relevant?
Draft a behavioral description in natural language.
Subsequently, translate this description into JML annotation.
You should only provide the JML annotation and not the Java code.
We will then use the KeY verification system to check if the JML annotation is correct.
If the program verification fails, you will be provided with information about the failure and asked to correct the JML annotation.
Always add the JML keyword `normal_behavior` to the contract -- this guarantees that no exceptions are being thrown.
Your answers should always have the following format where the the <X> is substituted by the JML annotation suggested by you:
<your natural language reasoning>
```
/*@ <X>
*/
```
\end{lstlisting}
Subsequently, we prompt GPT with the following prompt to
generate a contract:
\begin{lstlisting}[language={}]
Given the following Java class:
'''
<partially annotated file>
'''
Please provide a JML annotation to the method '<called method>' such that the contract specified by '<calling method>' is satisfied.
A contract for a submethod typically has the following structure:
```
/*@ normal_behavior // Ensures that no exceptions are thrown by submethod
  @ requires ...; // Requirements for an invocation of the submethod, based on the calling method.
  @ ensures ...; // Guarantees that the submethod will satisfy after the invocation.
  @ assignable ...; // Fields that are assignable in the called method.
@*/
```
\end{lstlisting}

For feedback-based approaches, we distinguish between syn-
tactical and semantical errors. For syntactical errors, we use
the following prompt to continue the conversation:
\begin{lstlisting}[language={}]
The provided code is not valid JML. Please try again and make sure to provide a valid JML contract.
This might describe the reason why change is required:
Beware that the error message may contain variable names from inside the proof system that are not available in the original code.
Such variable names must not in your JML answer; only use variable names from the original code.
<parser error>

Was expecting one of: <list of possible options>

Always add the JML keyword `normal_behavior` to the contract -- this guarantees that no exceptions are being thrown.
\end{lstlisting}
An example for a parser error message could be:
\begin{lstlisting}[language={}]
Error during JML parsing: Failed to parse JML fragment: Encountered unexpected token: \"loop_invariant\" \"loop_invariant\"
    at line 9, column 5.
\end{lstlisting}

For semantical errors we use the following prompt:
\begin{lstlisting}[language={}]
The provided JML does not solve the task for the method '<violated contract method>'.
The KeY verification told me that the JML is wrong because some proof goals were not closed:
Beware that the error message may contain variable names from inside the proof system that are not available in the original code.
Such variable names must not in your JML answer; only use variable names from the original code.
During verification, the following proof branches could not be closed:
<proof branch labels>
Please fix the JML contract.
Always add the JML keyword `normal_behavior` to the contract -- this guarantees that no exceptions are being thrown.
\end{lstlisting}

Examples for a proof branch label could be:
\begin{lstlisting}[language={}]
Post (<called method name>)
\end{lstlisting}

\end{document}